\documentclass[12pt,a4j,notitlepage,fleqn]{article}

\usepackage{setspace}
\usepackage[margin=1in]{geometry}
\usepackage[symbol]{footmisc}

\usepackage[T1]{fontenc}
\usepackage[utf8]{inputenc}
\usepackage{authblk}

\makeatletter

\newcommand{\fboxsubsec}[1]{
	\begin{flushleft}
		#1
	\end{flushleft}
	}
\renewcommand{\subsection}{\@startsection{subsection}{2}{0pt}
	{1ex}
	{0.5ex}
	{\reset@font\it\fboxsubsec}
	}
\makeatother

\title{The Evolution of Stock Market Efficiency in the US:\\
A Non-Bayesian Time-Varying Model Approach}%

\author{Mikio Ito$^{a}$, \ Akihiko Noda$^{b}$\thanks{\scriptsize Corresponding Author. E-mail: noda@cc.kyoto-su.ac.jp, Tel. +81-75-705-1510, Fax. +81-75-705-3227.} \ and \ Tatsuma Wada$^{c}$

{\scriptsize ${}^{a}$ \it Faculty of Economics, Keio University, 2-15-45 Mita, Minato-ku, Tokyo 108-8345, Japan}

{\scriptsize ${}^{b}$ \it Faculty of Economics, Kyoto Sangyo University, Motoyama, Kamigamo, Kita-ku, Kyoto 603-8555, Japan}

{\scriptsize ${}^{c}$ \it Faculty of Policy Management, Keio University, 5322 Endo, Fujisawa, Kanagawa 252-0882, Japan}}

\date{\empty}

\renewcommand\thefootnote{\arabic{footnote}}

\usepackage[dvips]{graphicx}

\usepackage[]{natbib}%
\usepackage{amsmath,amssymb}%
\usepackage{ascmac}%
\usepackage{multirow}%
\usepackage{lscape}%
\usepackage{subfigmat}

\usepackage{pifont}%
\usepackage{arydshln}%
\usepackage[format=hang]{caption}
\usepackage[all]{xy}
\usepackage{url}
\bibpunct{(}{)}{;}{a}{}{,}

\def\hsymbu#1{\smash{\lower1.7ex\hbox{\huge$#1$}}}

\def\ve #1{{\mbox{\boldmath $#1$}}}

\newcommand{\citetapos}[1]{\citeauthor{#1}'s \citeyearpar{#1}}

\def\ve #1{{\mbox{\boldmath $#1$}}}

\begin{document}

\begin{titlepage}

\renewcommand{\thepage}{}
\renewcommand{\thefootnote}{\fnsymbol{footnote}}

\maketitle

\vspace{-10mm}

\noindent
\hrulefill

\noindent
{\bfseries Abstract:} A non-Bayesian time-varying model is developed by introducing the concept of the degree of market efficiency that varies over time. This model may be seen as a reflection of the idea that continuous technological progress alters the trading environment over time. With new methodologies and a new measure of the degree of market efficiency, we examine whether the US stock market evolves over time. In particular, a time-varying autoregressive (TV-AR) model is employed. Our main findings are: (i) the US stock market has evolved over time and the degree of market efficiency has cyclical fluctuations with a considerably long periodicity, from 30 to 40 years; and (ii) the US stock market has been efficient with the exception of four times in our sample period: during the long-recession of 1873-1879; the recession of 1902-1904; the New Deal era; and the recession of 1957-1958 and soon after it. It is then shown that our results are partly consistent with the view of behavioral finance.\\



\noindent
\hrulefill

\end{titlepage}

\bibliographystyle{apecon}



\pagebreak


\section{Introduction}\label{sec:IT}

Over the past forty years, research on the efficient market hypothesis (EMH) has been advanced by a large number of researchers. If a market is ``efficient,'' then the information that is necessary to form rational expectations about future stock prices will be immediately transmitted to all market participants. As a result, no unexploited excess profit opportunities would be left in the market. In practice, however, testing EMH involves a difficult issue, which is a choice of an information set. Because information that market participants utilize is not clear in scope, econometric tests require us to choose the variables that represent the information set for the market participants. In this paper, we particularly focus on what the literature calls the weak-form of efficiency, which states that future returns can be predicted only by the information obtained from past returns.  

There are a large volume of studies of the weak-form of EMH, as clearly presented in a recent comprehensive survey by \citet{lim2011esm} for the empirical work on the weak-form of market efficiency. Utilizing autocorrelation tests and reviewing the preceding studies, \citetapos{fama1970ecm} seminal paper concludes that stock markets are almost always efficient.\footnote{His survey includes \citet{fama1965bsm}, \citet{fama1966fra}, \citet{fama1969asp}, \citet{jensen1968tpm}, and \citet{blume1970pta} among as others.} Two decades later, \citet{fama1991ecm} addresses the same issue shedding light on slightly different aspects such as anomalies arising from return seasonality (\citet{ariel1987ame,ariel1990hsr}, \citet{harris1986atd}, and \citet{keim1983sra,keim1989tpb}) and the firm size effect (\citet{banz1981trr}, \citet{conrad1988tve}, \citet{chan1991src}, and \citet{fama1992tcs}). After evaluating the vast literature concerning market efficiency, \citet{fama1991ecm} presents mounting evidence for the predictability of stock returns, indicating that markets are inefficient. Further, \citet{fama1998eml} again reviews the empirical work on event studies concerning several long-term return anomalies. 

With rigorous statistics, \citet{malkiel2003emh} and \citet{schwert2003ame} critically examine the validity of the predictability reported in stock returns. While the former generally agrees with \citetapos{fama1970ecm} conclusion, the latter reports a number of anomalies, concurring with \citetapos{fama1991ecm} conclusion. More recently, \citet{yen2008wmh} suggest that whether or not the EMH is supported depends upon sample periods: 1960s data are generally affirmative. Yet in the 1990s, this idea received attacks from the school of behavioral finance.

As \citet{malkiel2005meb} clearly point out, the controversy over the EMH between proponents of the EMH and advocates of behavioral finance is ongoing. Whereas criticisms of EMH from behavioral finance that assume limited rationality are quite strong, the lack of consensus on the EMH is partly attributable to the fact that previous studies have primarily focused on whether the returns of stock follow a random walk process. While the random walk returns support the EMH, returns following a non-random walk process do not necessarily rule out the EMH. In fact, as \citet{nyblom1989tcp} shows, a process with a single break point can also be a martingale process. Therefore, a rejection of the random walk hypothesis as null may present little information about the efficiency of the stock market.
However, there is an attempt to reconcile opposing sides. Considering an evolutionary alternative to market efficiency, \citet{lo2004amh,lo2005rem} proposes a hypothesis, which he calls the adaptive market hypothesis (AMH). In this new hypothesis, reconciliation is reached in a consistent manner between the proponents of the EMH and behavioral finance: market efficiency is not an object that is statistically tested, but a framework from which most researchers can discuss various perspectives. This framework reinforces the view that the stock market evolves over time and that market efficiency also varies with time. As \citet{lim2011esm} summarize, by allowing market efficiency to evolve over time, we are able to take into account a variety of factors that play different roles in the stock market. Those considered are: market participants' spontaneous irrational behavior and their mistakes-learning process, both of which are consistent with the main principles that advocates of behavioral finance espouse. Also considered are technological advances in the trading environment, such as rapid transmission of information and swift transaction of funds, among others. For example, \citet{gu2002eme} use daily data for the US stock market (Dow Jones Industrial Average) from 1896 to 1998. After computing autocorrelation for stock returns, they find that the US market evolves and that the market is efficient for the period of 1980 to 1998 due to technological advances and investors experience. \citet{emerson1997eme} examine the weak-form of market efficiency by adopting a model that has time-varying autoregressive parameters with an error term following a generalized autoregressive conditional heteroskedasticity (GARCH) process. The parameters are estimated via the Kalman filter and their model is applied to Bulgarian data. \citet{zalewska1999efs} also utilize the time-varying autoregressive plus GARCH error model and propose a new test for time-varying parameters within their framework. In line with this time-varying parameter approach, recent papers, such as \citet{ito2009mdt}, \citet{kim2011srp} and \citet{lim2013aus}, conclude that market efficiency (or the degree thereof) varies with time. There are still at least two questions left unanswered. First, is there any measure that quantifies the time-varying degree of market efficiency together with its statistical inference? If we have such a measure and its statistical inference, we will be able to say when an inefficient market shows up. Second, what causes the market inefficiency, which appears from time to time? Is it possible to interpret market inefficiency in a way that is consistent with the view of behavioral finance?

Given the remaining questions mentioned above, the contribution of this paper is twofold. First, new convenient methodologies are developed to examine the evolution of the US stock market. As is clearly explained in the following sections, our new measure of market efficiency has several advantages compared to those previously proposed. In particular, it only requires the underlying model to be approximated by an autoregressive model with time-varying coefficients. This feature is particularly appealing because we do not know the true underlying model that governs the behavior of stock prices. Also, the proposed technique is found to be very easy to implement; and applicable to a variety of models with time-varying coefficients. Second, with the concept and measure of the degree of market efficiency, it is shown that the US stock market changes over time and its degree of efficiency changes accordingly. In particular, we consider a non-Bayesian time-varying autoregressive (TV-AR) model, and apply it to the time-varying moving average (TV-MA) model. Then, our main results are the following: (i) the US stock market evolves (or fluctuates in terms of its structure regarding return predictability) over time and its market efficiency changes accordingly. More specifically, the degree of market efficiency in the US stock market has a cyclical fluctuation with a very long periodicity, from 30 to 40 years. (ii) the US stock market is mostly efficient, except for four times (three recessions and the New Deal era) in our sample period. Finally, with careful assessment, we are able to attribute the periods of market inefficiency to sudden changes in individuals' behavior or the irrationality of the individuals, which is compatible with the argument of behavioral finance. 

This paper is organized as follows. In Section 2, the model and new methodologies for our non-Bayesian TV-AR model are presented. The data on the US stock market, together with preliminary unit root test results, are described in Section 3. In Section 4, we show empirical results that the market efficiency in the US stock market periodically varies over time and that the amplitude of the cycle decreases in the long-run. Section 5 concludes. The online appendix provides mathematical and statistical discussions about our new estimation methodologies.\footnote{Online appendix is available at \url{http://at-noda.com/appendix/evolution_appendix.pdf}}

\section{Model and Methodology}\label{sec:MM}
\subsection{Preliminaries}

For a representative household, the first order conditions for its utility maximization problem result in the following Euler equation (see, for example, \cite{cochrane2005ap}): 
\begin{equation}
p_{t}=E_{t}\left[ m_{t+1}\left( p_{t+1}+\kappa _{t+1}\right) \right] ,
\label{Euler}
\end{equation}%
where $p_{t}$ is the stock price; $m_{t+1}$ is the stochastic discount
factor;\footnote{The stochastic discount factor is defined as 
\begin{equation*}
m_{t+1}=\delta \frac{u^{\prime }\left( C_{t+1}\right) }{u^{\prime }\left(
C_{t}\right) },
\end{equation*}%
where $\delta $ is the subjective discount factor; $u^{\prime }\left(\cdot \right) $ is the first derivative of a utility function; and $C_{t}$ is consumption at $t$.} $\kappa_{t+1}$ is the dividend; and $E_{t}\left[ \cdot \right] $ represents the conditional expectation given the information available at $t$. Provided $m_{t+1}$ is constant over time and close to 1,%
\footnote{If the household has a risk-neutral preference, 
\begin{equation*}
m_{t+1}=\delta 
\end{equation*}%
for all $t$.} together with the condition $E_{t}\left[ \kappa_{t+1}\right]=0$, the stock price should follow a random walk process (or a martingale). In such a case, the expected return is unpredictable since the expected (gross) return is always constant and one, or $E_{t}\left[p_{t+1}/p_{t}\right] =1$. As \citet{fama1970ecm} states, any test regarding equation (\ref{Euler}) -- including unit-root tests of the stock price -- is essentially a test of a joint null hypothesis of market efficiency and the model of the market equilibrium.

Therefore, it is possible that equation (\ref{Euler}) does not hold in times when: i) the household is risk averse and its stochastic discount factor ($m_{t+1}$; the marginal rate of substitution of two periods) varies; ii) $E_{t}\left[\kappa_{t+1}\right] \neq 0$; or iii) the market is not efficient (not in equilibrium). While this paper investigates whether equation (\ref{Euler}) holds, our approach here is to find the time periods for which equation (\ref{Euler}) is true, rather than to test a unit root in $p_{t}$ for the entire sample period. In doing so, we allow $p_{t}$ to follow a very flexible process (namely, the time-varying autoregressive process) that permits the autoregressive coefficients to change from time to time. The reason why we choose this approach is that we are more interested in the time-varying nature of the market, or the evolution of the market, which is perhaps captured by shifts or fluctuations in the parameters or variables that pertain to equation (\ref{Euler}).

Our main idea stems from the observation in Figure \ref{ev_fig1}, which exhibits returns on the S\&P 500 from January 1871 through June 2012. It is quite evident that the Wall Street Crash in October 1929 and subsequent years, known as the Great Depression, caused an irregular pattern in returns. Due to the fact that outliers often lead to incorrect conclusions in statistical testing (see for example \citet{perron1989tgc}), one way to carefully investigate whether the efficiency condition (\ref{Euler}) is satisfied is to exclude the Great Depression period from our sample. Still, a practical question remains: Which observation(s) should be removed from our sample as a result of the Great Depression? Without getting into data mining, we employ a model that allows flexible specifications for the process of the variable, namely, a time-varying coefficients model.

\subsection{Impulse-Responses and The Long-Run Multiplier}

From the argument in the previous subsection, our main focus is reduced to the condition 
\begin{equation}
E\left[ x_{t} \mid \mathcal{I}_{t-1}\right] =0  \label{EMH}
\end{equation}%
where $x_{t}=\ln p_{t}-\ln p_{t-1}$. In other words, the time-$t$ expected capital gain given the information set available at $t-1$ is zero.

Assuming that $x_{t}$ is stationary, by the Wold decomposition we write the time-series process of $x_{t}$ as 
\begin{equation*}
x_{t}=\phi \left( L\right) u_{t},
\end{equation*}%
where $\phi \left( L\right) =\phi _{0}+\phi _{1}L+\phi _{2}L^{2}+\cdots $,%
\footnote{We also assume $\sum_{i=0}^{\infty }\phi _{i}^{2}<\infty $.} with $\phi_{0}=1$; $L$ is the lag operator; $\left\{ u_{t}\right\} $ is an i.i.d. process with the mean of zero, and a variance of $\sigma ^{2}$. Note that equation (\ref{EMH}) holds if and only if $\phi \left( L\right) =1$. Put differently, under the assumption that the discount factor is constant and close (or equal) to 1, EMH is equivalent to $x_{t}=u_{t}$. That is, the capital gain is random and serially uncorrelated, so an unanticipated shock at $t$ affects the capital gain at $t$, but not in any subsequent periods. This argument leads us to two important features of the capital gain under EMH: First, the long-run multiplier of a shock, i.e., 
\begin{equation*}
\phi _{\infty }\equiv \phi \left( 1\right) =\phi _{0}+\phi _{1}+\phi
_{2}+\cdots ,
\end{equation*}%
is always 1.\footnote{There are at least two reasons why $\phi(1)$ is called the long-run multiplier. The first one is that it measures the long-run effect of a shock $(u)$ on the log of the stock price, $(p)$. The second is that the spectral density of stock returns $(x)$ at the frequency $0$, which is sometimes called the long-run variance, is $\phi(1)^2$ times a constant $(\sigma^2/2\pi)$. Also, it should be noticed that $\phi\left(L\right)=1$ implies $\phi\left(1\right)=1$, while the opposite is not necessarily true.} Second, impulse-responses quickly disappear after impact. In other words, the effect of an unanticipated shock on the capital gain is short-lived. Therefore, in this paper, we shall compute both the long-run multiplier and the impulse-response functions of a shock in order to investigate whether the EMH holds.

\subsection{The Non-Bayesian Time-Varying AR Model}\label{sec:TV-AR}

Assuming process $x$ is invertible, we estimate the following time varying AR($q$) model
\begin{equation}
x_{t}=\alpha_{0}+\alpha_{1,t}x_{t-1}+\alpha_{2,t}x_{t-2}+\cdots+\alpha_{q,t}x_{t-q}+\varepsilon_{t},  \label{tv_arq}
\end{equation}
where $\varepsilon_{t}$ is an error term with $E\left[\varepsilon_{t}\right]=0$, \ $E\left[\varepsilon_{t}^{2}\right]=\sigma^{2}$, and $E\left[\varepsilon_{t}\varepsilon_{t-m}\right]=0$ for all $m\neq 0$.

The martingale formulation -- introduced by \citet{nyblom1989tcp} and \citet{hansen1992a} -- has substantial flexibility and covers a wide range of parameter dynamics. For instance, it allows for a stochastic process with a single structural break (as \citet{nyblom1989tcp} points out) and a random walk (as specified in many articles). \citet{nyblom1989tcp} and \citet{hansen1992a} provide several test statistics for the null hypothesis of the constant parameters against the alternative hypothesis of at least one martingale parameter, in both linear and non-linear models.

We use \citetapos{hansen1992a} statistic in this paper.\footnote{For the detail of test statistics, see \citet{hansen1992a}.} As for the alternative hypothesis, we consider a model in which all the AR coefficients, except for the one that corresponds to the intercept term, follow independent random walk processes. More specifically, under the alternative hypothesis, we have  
\begin{equation}
\alpha_{l,t}=\alpha _{l,t-1}+v_{l,t} \ \ (l=1,2,\cdots,q)\label{rw_a}
\end{equation}
where $\left\{v_{l,t}\right\}$ satisfies $E\left[v_{l,t}\right]=0$, \ $E\left[v_{l,t}^{2}\right]=\sigma^{2}$ and $E\left[v_{l,t}v_{l,t-m}\right]=0$ for all $l$ and $m\neq 0$. This is because, as we will see in Section 4, the data in the U.S. stock market are in favor of unstable autoregressive coefficients. 

If, instead, the AR coefficients were constant over time (i.e., an ordinary AR model), one could estimate them using the following linear regression model:\footnote{One could estimate this model by the ordinary least squares (OLS), which is used in this paper. Because of this, our estimation method can be seen as conditional maximum likelihood estimation, see \citet{hamilton1994tsa}.}

\begin{equation}
\ve{x}=\left( 
\begin{array}{cc}
1 & {}^{t}\ve{z}_{1} \\ 
1 & {}^{t}\ve{z}_{2} \\ 
\vdots & \vdots \\ 
1 & {}^{t}\ve{z}_{T}%
\end{array}%
\right) \ve{\alpha}+\ve{u} \label{bold_x}
\end{equation}%
where $\ve{x}={}^{t}(x_{1} \ x_{2} \ \cdots \ x_{T})$, $\ve{\alpha}={}^{t}(\alpha_{0} \ \alpha_{1} \ \cdots \ \alpha_{q})$, $\ve{u}={}^{t}(u_{1} \ u_{2} \ \cdots \ u_{T})$, $\ve{z}_{t}={}^{t}(x_{t-1} \ x_{t-2} \ \cdots \ x_{t-q})$; superscript $t$ denotes a transpose of a vector.

Alternatively, when a non-Bayesian time-varying model with AR coefficients that evolve by random walk processes (specified in the above martingale formulation; equations (\ref{tv_arq}) and (\ref{rw_a})) is estimated, (\ref{bold_x}) should be modified; and our model can be set in a state space form:
\begin{eqnarray}
x_{t} &=&\alpha_{0}+{}^{t}\ve{z}_{t}\widetilde{\ve{\alpha}}_{t}+u_{t}%
 \ \ (t=1,2,\ldots,T) \label{obs} \\
\widetilde{\ve{\alpha}}_{t} &=&\widetilde{\ve{\alpha}}_{t-1}+v_{t}%
 \ \ (t=1,2,\ldots,T) \label{stat}
\end{eqnarray}%
where $\widetilde{\ve{\alpha}}_{t}={}^{t}(\alpha _{1,t} \ \alpha _{2,t} \ \cdots \ \alpha _{q,t})$. Our model is non-Bayesian because it does not necessitate the prior distributions of parameters.
 
Equations (\ref{obs}) and (\ref{stat}) are called the observation equation and the state equation, respectively. Notice that we assume the intercept $\alpha_{0}$ is time invariant. This is to avoid over-fitting that would occur if the time-varying intercept were employed.\footnote{Another reason why we assume a constant intercept is that the unit root tests fail to reject the null hypothesis that $x$ has a unit root. Allowing the intercept to vary with time could make $x$ non-stationary. Because our model has the time-varying AR coefficients, this assumption does not preclude variations in the ex-ante equity return unless $x$ is a white noise process.}

Similar to the idea that is illustrated in \citet[pp.469--470]{maddala1998urc} and the method employed in \citet{ito2009mdt}, we regard equations (\ref{obs}) and (\ref{stat}) as a system of simultaneous equations:\footnote{While this method and the method explained in \citet{maddala1998urc} are seemingly alike, this paper proposes an efficient (non-Bayesian) estimation of time-varying AR models, whereas the latter deals with a general time-varying coefficient model with exogenous regressors (i.e., without an AR component). With regard to \citet{ito2009mdt}, this paper formalizes their method in that ours provides not only efficient estimation procedures but also statistical inferences for the estimated time-varying parameters. In fact, \citet{ito2007nme}, which is an unpublished manuscript cited in \citet{ito2009mdt}, has subsequently become the appendix of this paper.}

\begin{eqnarray}
\ve{x} &=&Z\ve{\beta}+\ve{u}  \label{eq_x} \label{Ito-obs}\\
\ve{\gamma} &=&W\ve{\beta}+\ve{v}  \label{eq_g} \label{Ito-state}
\end{eqnarray}%
where 
\begin{equation*}
\underbrace{Z}_{T\times \left(1+qT\right)}=\left( 
\begin{array}{cccc}
1 & {}^{t}\ve{z}_{1} &  & \mathit{O} \\ 
\vdots &  & \ddots &  \\ 
1 & \mathit{O} &  & {}^{t}\ve{z}_{T}%
\end{array}%
\right) ; \ \ \underbrace{W}_{qT\times \left( 1+qT\right)}=\left( 
\begin{array}{ccccc}
0 & -I &  &  & \mathit{O} \\ 
0 & I & -I &  &  \\ 
\vdots &  & \ddots & \ddots &  \\ 
0 & \mathit{O} &  & I & -I%
\end{array}%
\right) ;
\end{equation*}%
$I$ is a $q\times q$ identity matrix; $\ve{\beta}={}^{t}\left(\alpha_{0} \ {}^{t}\widetilde{\ve{\alpha}}_{1} \ \cdots \ {}^{t}\widetilde{\ve{\alpha}}_{T}\right)$; $\ve{\gamma}={}^{t}\left(-{}^{t}\widetilde{\ve{\alpha}}_{0} \ {}^{t}\ve{0} \ \cdots \ {}^{t}\ve{0}\right)$; and $\ve{v}={}^{t}\left(v_{1,1} \ \cdots \ v_{q,1} \mid v_{1,2} \ \cdots \ v_{q,2} \mid \cdots \ \cdots \ \cdots \mid v_{1,T} \ \cdots \ v_{q,T}\right)$.

Note that one can regard $\widetilde{\ve\alpha}_{0}=$ $^{t}\left(\alpha_{1,0} \ \alpha_{2,0} \ \cdots \ \alpha_{q,0} \right)$ as a prior vector of non-Bayesian TV-AR coefficients, while $\alpha_{0}$ in the observation equation (\ref{obs}) represents the constant intercept term. For convenience, we stack equations (\ref{eq_x}) and (\ref{eq_g}), in the system of simultaneous equations.
\begin{equation}
\left[ 
\begin{array}{c}
\ve{x} \\ 
\cdots \\ 
\ve{\gamma }%
\end{array}%
\right] =\left[ 
\begin{array}{c}
Z \\ 
\cdots \\ 
W%
\end{array}%
\right] \ve{\beta }+\left[ 
\begin{array}{c}
\ve{u} \\ 
\cdots \\ 
\ve{v}%
\end{array}%
\right]  \label{sys_eq}
\end{equation}
We estimate the non-Bayesian TV-AR coefficients by applying least squares techniques, OLS or generalized least squares (GLS), to equation (\ref{sys_eq}).\footnote{In the online appendix, we present a least square estimator of a state space model that involves the usual Kalman smoother. Also, the online appendix formally discusses why the least squares technique can be used to estimate the non-Bayesian TV-AR coefficients. While the regressor of our system in equation (\ref{sys_eq}) may be an extremely large matrix of the data exceeding one thousand, this does not pose much of a problem for most recent computers. Indeed, in our case, only several minutes are necessary for estimation of the parameters, impulse responses, and the long-run multiplier. \citet{paige1977lse} show a numerically efficient method for this framework.}

The three major advantages of our method over the conventional Kalman smoothing (e.g., \citet{hamilton1994tsa}) are as follows.

First, our method is quite simple and fast. Unlike the conventional Kalman filtering and following smoothing, no iteration is required. Second, a wide variety of models can be easily dealt with even when the state equations of such models are not represented by simple stochastic difference equations. This is especially beneficial for models with random parameter variations because stochastic constraints or moment conditions can simply be put in the state equation. Third, it is evident that our estimation is based on classical regression. Hence, a number of (asymptotic) properties regarding estimates are preserved. In addition, in the case of non-Normal errors, a correcting method, such as GLS, can be readily applied. In particular, non-Normal errors are allowed for both the observation errors $\ve{u}$ \textit{and} the state equation errors $\ve{v}$: not only heteroskedasticity in errors ($\ve{u}$, $\ve{v}$, or both), but also correlation between observation errors and state equation errors (correlation between $\ve{u}$ and $\ve{v}$) is permitted. In such cases, covariance estimators, for example, \citet{newey1987sps,newey1994als} can be utilized for GLS. The least squares method for (\ref{sys_eq}) gives us:

\begin{equation*}
\widehat{\ve{\beta}}=\underset{\left(1+qT\right) \times \left(
T+qT\right) }{Q}\left[ 
\begin{array}{c}
\ve{x} \\ 
\cdots  \\ 
\ve{\gamma} 
\end{array}%
\right] 
\end{equation*}%
where $Q$ is a $\left( 1+qT\right) \times \left( T+qT\right) $ matrix.

One of the interesting features of our method is that the smoothed estimate of the time-varying AR coefficient vector alpha can be represented by the weighted average of the observed data and a constant.\footnote{This is due to the fact that the smoothed estimator is, in its probability limit, the projection of the state vector onto the space spanned by the data and the constant. See \citet[Chapter 13]{hamilton1994tsa} and \citet[Chapter 10]{sargent1987mt}, among others. The relation of our approach to the moving-window method is described in the online appendix.}

\subsection{Time-Varying Impulse Responses and The Time-Varying Long-run Multiplier (Degree of Market Efficiency)}\label{subsec: 2.4}

In this subsection, we present the method that provides time-varying impulse responses and time-varying long-run multiplier. They are calculated from the non-Bayesian TV-AR coefficients in each period utilizing the method described in previous subsections. Statistical inference on our estimates can be conducted by the Monte Carlo technique under the hypothesis that all of the TV-AR coefficients are zero.

While this idea is quite simple, the following two caveats need special attention: (1) the non-Bayesian TV-AR model that we estimate is only an approximation of the real data generating process, which may be a very complex process; and (2) we consider the estimated stationary AR($q$) model index by each period $t$ as a local approximation of the underlying complex process.

First, let us consider a time-varying AR($q$) model. To find the appropriate order of $q$, we employ the Schwartz Bayesian information criterion (SBIC). After the order, $q$, is selected, we estimate the non-Bayesian TV-AR($q$) by the method presented in Section 2.3.

Second, from the non-Bayesian TV-AR($q$) model, the TV-MA($\infty$) model is derived:
\begin{equation*}
x_{t}=u_{t}+\phi_{1,t}u_{t-1}+\phi_{2,t}u_{t-2}+\cdots.
\end{equation*}
The coefficients of the TV-MA($\infty$) can be readily computed from the estimated TV-AR($q$) in the following manner: With the estimates ${}^{t}\left(\widehat{\alpha}_{1,t},\ldots,\widehat{\alpha}_{q,t}\right)$ and $\widehat{\alpha}_{0}$, and Proposition 2.4 of \citet[Section 2.3.2]{lutkepohl2005nim}, the TV-MA($\infty$) coefficients are found by using:
\begin{equation*}
\widehat{\phi}_{0,t}=1, \ \ \ \widehat{\phi}_{k,t}=\sum_{j=1}^{k}%
\widehat{\phi}_{k-j,t}\widehat{\alpha}_{j,t}.
\end{equation*}
The time-varying long-run multiplier associated with the time-varying coefficients can be computed by, (see equation (2.3.26) in \citet[p.56]{lutkepohl2005nim}), 
\begin{equation}\label{LongRunMult}
\widehat{\phi}_{\infty,t}=\frac{1}{1-\widehat{\alpha}_{1,t}-\widehat{\alpha}_{2,t}-\cdots-\widehat{\alpha}_{q,t}}.
\end{equation}
We pay special attention to the time-varying long-run multiplier because it measures the deviation from efficient market. Note that in the case of efficient market where $\phi_1=\phi_2=\cdots=0$ and $\alpha_1=\alpha_2=\cdots=\alpha_q=0$, the long-run multiplier $\phi_{\infty,t}$ becomes one; otherwise, $\phi_{\infty,t}$ deviates from one. Hence, we call $\phi_{\infty,t}$ the degree of market efficiency; and we consider the large deviations of $\phi_{\infty,t}$ from 1 (both positive and negative) to be evidence of market inefficiency.

Note, however, that our measure being 1 does not necessarily mean the process exhibits no autocorrelation, nor is it a sufficient condition for equation (\ref{EMH}) being true; although the opposite is always true. What the degree of market efficiency really measures is the power spectrum of process $x$ at frequency zero; or, the long-run impact of a shock. Conceptually, the power spectrum at frequency zero measures the permanent effect of the shock to the process (more precisely, the relative variance of the component of $x$ that has a periodicity of infinity). By using this measure, together with impulse-responses, we can determine whether process $x$ bears a resemblance to a white noise process, which clearly satisfies (\ref{EMH}). If equation (\ref{EMH}) does not hold, and $x$ has strong persistence, then the degree is higher than 1 and the power spectrum at zero frequency should be much larger than 1 (times $\sigma^{2}/2\pi$)\footnote{Alternatively, one can use a different degree of market efficiency that involves the absolute values of the estimated AR parameters. Yet, we have some reservations in regard to this statistic because we are uncertain about the inference of this measure. We are also uncertain as to whether this measure converges in distribution since it is not a continuous function of MLE. Hence, we do not employ this statistic.}.

In order to conduct statistical inference on our time-varying impulse response, we build a set of Monte Carlo samples of the TV-AR estimates under the hypothesis that all of the TV-AR coefficients are zero. That is, we derive a (simulated) distribution of the estimated TV-AR coefficients assuming that the stock return processes are generated under the efficient market hypothesis. In addition, we can compute the corresponding distributions of the impulse response and long-run multiplier based on the simulated one. Finally, we conduct statistical inference on our estimates by using confidence bands derived from such simulated distributions.

\section{Data}\label{sec:DAT}

Monthly returns for the S\&P500 stock price index from January 1871 through December 2012 (obtained from Robert Shiller's website) are utilized. In practice, we compute the first-difference of the logarithm of the S\&P500 stock price index as the returns. Figure \ref{ev_fig1} presents time series plots of the returns for the S\&P500.
\begin{center}
(Figure \ref{ev_fig1} here)
\end{center}

For the purpose of estimation, any variable that appears in the moment conditions should be stationary. To check whether the variables satisfy the stationarity condition, we apply the ADF-GLS test of \citet{elliott1996eta}. When the procedure proposed by \citet{ng2001lls} is utilized, this unit root test is robust against size-distortions. The results of the ADF-GLS test along with descriptive statistics of the data are presented in Table \ref{ev_table1}: The ADF-GLS test rejects the null hypothesis that the variable contains a unit root at conventional significance levels.\footnote{For selecting the optimal lag length, we employ the Modified Bayesian Information Criterion (MBIC) instead of the Modified Akaike Information Criterion (MAIC). This is because, from the estimated coefficient of the detrended series, $\widehat{\psi}$, we do not find the possibility of size-distortions (see \citet{elliott1996eta}; \citet{ng2001lls}).}
\begin{center}
(Table \ref{ev_table1} here)
\end{center}

\section{Empirical Results}\label{sec:ER}

In this section, we report three sets of results. They are: i) our preliminary estimation together with \citetapos{hansen1992a} test which confirm that the parameters in the standard AR model are not constant over the sample period; ii) estimation of the non-Bayesian TV-AR model which reveals the validity of our model; and iii) the impulse-responses and the long-run multiplier that suggest market efficiency for a limited period of time.

\subsection{Preliminary Estimation and Parameter Constancy Test}

Assuming a standard AR($q$) model with constant parameters, we utilize the SBIC of \citet{schwarz1978edm} to select the lag-order, $q$. As a result, $q=2$ -- the second order autoregressive model -- is obtained.\footnote{The heteroskedasticity and autocorrelation consistent (HAC) covariance matrix estimator of \citet{newey1987sps,newey1994als} is used.}
\begin{center}
(Table \ref{ev_table2} here)
\end{center}
Our estimation result for an AR(2) model with the whole sample is summarized in Table \ref{ev_table2}: all AR estimates are statistically significant at the 1\% level. It is notable that the first-order autoregressive estimate is about $0.31$ (and the second one is about $-0.08$). This implies that approximately 10\% of an unanticipated shock to the average stock return will remain in the average stock returns two months later.

Are the AR coefficients constant over the sample period? One approach that we consider useful is to apply a test of the parameter constancy. As presented in Table \ref{ev_table2} (the entry below $L_{C}$), \citetapos{hansen1992a} test rejects the null hypothesis of constant parameters at the 1\% significance level (The asymptotic critical value at the 1\% significance level is 1.60). Having found non-constant parameters in the AR($q$) model, we move forward to focus on the time-varying AR model in order to see whether gradual changes occur in the US stock market.

\subsection{Non-Bayesian TV-AR Estimation}

Given the fact that the test of the parameter constancy rejects the null hypothesis against the alternative hypothesis of the AR parameters following the random walk process, our non-Bayesian time-varying estimation method is carried out to estimate our TV-AR(2) model. Because of the properties we discussed in the online appendix, our method is shown to have a particular advantage over the simple moving window method that assumes a fixed width to compute AR coefficients or the correlation coefficient.\footnote{This point distinguishes our work from previous studies, such as \citet{kim2011srp} and \citet{lim2013aus}.} With optimally selected window widths, the coefficients of the TV-AR(2) model are computed.\footnote{While our method for parameter estimation does not require the restriction that the AR parameters are locally stationary, we confirm ex-post that our model is locally stationary -- the stationary AR(2) process for all $t$ -- by checking the roots of the equation $1-\alpha_{1,t}z-\alpha _{2,t}z^{2}=0$. All of the roots lie outside the unit circle.}
\begin{center}
(Figure \ref{ev_fig2} here)
\end{center}
Figure \ref{ev_fig2} presents the weight for $t=851$ (February, 1942). As is discussed in Section 2.4, the estimate utilizes a wide range of observations. In this case, the smoothed estimate for $t=851$ requires the data points of 170 months.
\begin{center}
(Figure \ref{ev_fig3} here)
\end{center}

Demonstrated in Figure \ref{ev_fig3} (solid lines), the estimated AR coefficients are very unstable over time.\footnote{Yet, \citetapos{ljung1978oml} test for the residuals do not reject the null hypothesis of autocorrelation. To estimate the TV-AR model, the HAC estimator of \citet{newey1987sps,newey1994als} is employed.} As a statistical inference, we provide significance bands in Figure \ref{ev_fig3}. Due to the fact that our method is based on least squares for subsamples (i.e., widths), our estimates may suffer from a downward bias (see for example, \citet{andrews1993emu}). Taking into account such a possibility, our significance bands for the estimates are constructed as follows. For the data generating process assuming $\alpha_{1,t}=\alpha_{2,t}=0$ for all $t$, the TV-AR(2) model is estimated. Repeating this process 5000 times and tabulating $\widehat{\alpha}_{1,t}$ and $\widehat{\alpha}_{2,t}$ for each $t=1,\ldots ,T$, we plot the 99\% upper and lower limits of the estimates. Therefore, Figure \ref{ev_fig3} shows that the AR(1) coefficient is significant most of the time, while the AR(2) coefficient is not.

From the estimated AR(1) coefficient, remarkably, the market crash and the following Great Depression did not cause a great deal of deviation of the AR(1) coefficient from its historical average: In fact, in the late 1980s, a slightly larger magnitude of deviation can be seen. However, care must be taken in interpreting the estimates of the AR coefficients. Our ultimate goal in this paper is to compute the long-run multiplier, defined in equation (\ref{LongRunMult}), and the time-varying impulse-responses to see whether there is evidence of market efficiency. The next subsection presents these two measures.

\subsection{Time-Varying Impulse Responses and The Time-Varying Degree of Market Efficiency}\label{subsec:TVIR}

Figure \ref{ev_fig4} exhibits time-varying impulse responses. 
\begin{center}
(Figure \ref{ev_fig4} here)
\end{center}
Once again, there are two main reasons why we present the time-varying impulse responses in this paper. The first reason is that a comparison between time-varying impulse responses at different time periods allows us to visually grasp how the US stock market changes the way it responds to an unanticipated shock (to stock returns). Second, the impulse response function itself gives us some idea as to whether the market is efficient. Because of the way we define the degree of market efficiency (\ref{LongRunMult}), the impulse response function for an efficient market should exhibit the effect of an unanticipated shock disappearing immediately (i.e., no effect remains one period after the shock hits the market). Instead, if the market is inefficient, the impulse response function takes a longer time to converge to zero. Thus, the time needed for the impulse response to converge to zero can be understood as the speed of adjustment (toward the efficient market). Note that a lengthy adjustment process at one point in time may be interpreted in favor of the behavioral finance's view that the market participants change their behavior, thereby creating an inefficient market at that point in time.

Notably, the time path of an exogenous shock's effect on return varies widely with time. For example, in December 1919, when the estimated time-varying AR(1) coefficient reaches its whole sample minimum, only less than 20\% of the shock to the average stock return remains two months after the shock (Figure \ref{ev_fig4}, bottom left); whereas, in November 1987, when the estimated AR(1) coefficient reaches its whole sample maximum, more than 40\% of the shock is preserved two months after impact (Figure \ref{ev_fig4}, bottom right).

Although impulse-response functions demonstrate the speed of adjustment graphically, they do not provide decisive information about whether the US stock market is efficient at any given time (In fact, November 1987 was NOT the time that the long-run multiplier suggests market inefficiency). Now, we present our time-varying long-run multiplier -- the degree of market efficiency -- in Figure \ref{ev_fig5}.
\begin{center}
(Figure \ref{ev_fig5} here)
\end{center}
There are two important observations to be put forward.

First, despite the extraordinary magnitudes of the Great Crash of 1929 and Financial Crisis of 2008 (as displayed in Figure \ref{ev_fig1}), the degrees of market efficiency for such periods are not shown to be outliers.

Second, after the 1930s, the degree of market efficiency tends to deviate from one (thus, the market becomes relatively inefficient) during periods of expansion, and tends to be close to one (thus, the market becomes relatively efficient) during periods of contraction. When the monthly growth rate of the Industrial Production (IP) Index from January 1930 to December 2012\footnote{A quarterly GDP series is only available after 1947.I.} is computed and compared with the degree of market efficiency, the correlation coefficient is weak (0.133) but statistically significant (i.e., we are able to reject the null hypothesis of no correlation between IP growth and the degree of market efficiency).   

With that being said, the degree of market efficiency is greater than 1 for the entire sample period. Does this make our argument invalid? In order to investigate the validity of the argument, first we provide significance bands for the degree of market efficiency.\footnote{The statistical inference for the time-varying degree of market efficiency can deal with the following issue. The estimator of the time-varying degree of market efficiency for the beginning and end of the sample, inevitably, does not utilize as many data points as that for the middle of the sample; thus, Figure \ref{ev_fig5} may display imprecise estimation at the beginning and end of the sample.}

By doing so, we arrive at the conclusion that the stock market is efficient for the most of our sample period at the 1\% significance level. With such bands, we are able to find at least four clearly inefficient markets in our sample. Of the four inefficient markets, three of them have their highest degree of market efficiency (i.e., the largest deviation from market efficiency) during recessions. The first one appears during the long-depression (1873-1879), following the financial panic of 1873. Note that this is the longest recession NBER has ever recorded (65 months). The second inefficiency take place during the recession of 1902-1904, which follows the panic of 1901. The third peak is seen in August 1958 when a short, but very severe recession just passes its ``trough.'' A study by \citet{perron2009ltb} reveals that the cyclical component of the US post war real GDP, after taking into account the structural break in the slope of the trend in 1973, reaches its lowest point in the 1957-1958 recession. This result is in accordance with the trend-cycle decomposition by the Beveridge-Nelson decomposition (\citet{beveridge1981nad}; see \citet{morely2003wbn}). Also, it is worth noting that we do not find evidence of market inefficiency after 1958. This could be due to technological advances and investors' experience, as argued by Gu and Finnerty (2002) who use the daily series of the Dow Jones Industrial Average from 1896 to 1998 and find that market efficiency is attained only after 1980.\footnote{They do not use the time-varying parameter model. The differences between their findings and ours, especially market efficiency in the period of oil price shocks (1970s), need further investigation.} One exception took place in the expansion between 1933 and 1937, the aftermath of the Great Depression; and the economy was recovering only due to the aggressive fiscal policy called the New Deal.

One finding stands out: Not all recessions (or expansions) create inefficient markets, although some turning points of business cycles and those of the degree of market efficiency seem to be related. Therefore, we can conclude that the degree of market efficiency does not fluctuate as often as macrovariables such as GDP and consumption. From the view point of the spectral analysis, Figure \ref{ev_fig6} confirms this conclusion.\footnote{The power spectrum is estimated by utilizing the Bartlett window and the Quadratic window in order for the estimate to be consistent (See, for example, \citet{brockwell1991tst}). To select the bandwidth, \citetapos{andrews1991hac} method that are designed to consistently estimate the spectral density function at frequency zero is employed. We also examine the consistent estimator proposed by \citet{newey1994als} that is also designed to estimate the zero-frequency spectral density consistently. From Figure \ref{ev_fig6}, all estimators of the spectral density exhibit qualitative similarity.}
\begin{center}
(Figure \ref{ev_fig6} here)
\end{center}
The power spectrum of the estimated degree of efficiency has a peak and its most power in lower frequencies than in standard business cycle frequencies (periodicity corresponding to 72-384 months; shaded in Figure \ref{ev_fig6}. See also \citet{baxter1999mbc}.), indicating that the degree of market efficiency has a very long periodicity. Thus, market inefficiency emerges only infrequently.

\subsection{What Caused Market Inefficiency?}
Our result that market inefficiency emerges during extraordinary times leads us to one of the central questions in the literature. What causes market inefficiency? In particular, do market participants become irrational and make a number of mistakes when such events occur? One possibility is that in the face of an unusually large magnitude of events, the market participants alter their stochastic discount factor ($m$ in equation (\ref{Euler})), which is the function of: the consumption growth rate, the degree of risk aversion, and the subjective discount factor.\footnote{Since our focus is on the long-run aspects of the US stock market and the US economy, it is most appropriate to assume the constant relative risk aversion (CRRA) utility function. This is due to the fact that the CRRA utility function, as presented by \citet{king1988pgb} show, yields models that are consistent with the long-run economic fluctuations observed in the data. With the CRRA utility function, the (log-linearized) stochastic discount factor is simply the growth rate of consumption multiplied by a constant.} In this case, rational expectations are still a valid assumption for an individual's behavior; therefore, equation (\ref{Euler}) holds as true, but testing EMH by examining the time-series properties of stock returns is not appropriate.

To investigate whether $m$ changed abruptly in the sample period, we compute the month to month percent changes in the IP index (January 1919 onward). The growth of the IP index is a proxy for the growth of real consumption, which is only available after 1947 as a quarterly series.\footnote{It is found that the annual IP growth rate is not a weak instrument for the annual real consumption growth rate, excluding durable goods, during 1929-2013. The F-statistic is 31.23, whereas the rule of thumb requires the F-statistic be greater than 10.} Our conjecture is that if the IP growth is highly correlated with the long-run multiplier and if they are moving in the same direction during the times of deviation periods, then we conclude that market inefficiency is caused by changes in $m$. But if not, then, $m$ is unlikely to be the reason for market inefficiency. Figure \ref{ev_fig7} reveals that IP growth does not coincide the degree of the market efficiency for the periods in which the degree of market inefficiency suggests that the market is inefficient (i.e., the New Deal era and 1957-1958). 
\begin{center}
(Figure \ref{ev_fig7} here)
\end{center}
Moreover, IP growth has virtually no correlation with the time-varying degree of market efficiency, even if up to 2 months of leads and lags are considered.\footnote{This is different from the result presented in Section \ref{subsec:TVIR}, where correlation is computed after January 1930.} This means that $m$ does not change dramatically unless discrete jumps in the subjective discount factor or the degree of risk averse occur.

The second possibility is that equation (\ref{Euler}) does not hold for the periods in which the US stock market is inefficient. This is a case where market participants become irrational at least temporarily, and make mistakes, supporting the main argument of behavioral finance.

All in all, what our results indicate is that the market inefficiency that appears in the US stock market is most likely attributable to sudden and abrupt changes in individual's behavior, including possible violation of the optimality condition (\ref{Euler}), when extraordinary events take place in the US economy.

\section{Concluding Remarks}\label{sec:CC}
Focusing on the weak form of market efficiency which may vary with time, we develop a non-Bayesian time-varying model to examine whether or not the U.S. stock market evolves over time. In particular, the non-Bayesian time-varying AR (TV-AR) model is applied by taking into account various possibilities, namely, structural changes, regime shifts, and gradual changes that arise from a number of factors such as technological progress, natural shock, and individual's responses to them. In addition, a new measure of the degree of market efficiency is introduced and estimated. With a new and convenient technique, it is found that the U.S. stock market changes slowly over time: Our estimated power spectrum indicates the periodicity of the degree of market efficiency is 30 to 40 years. After careful consideration based on statistical inferences, the degree of market efficiency is found to be in favor of efficient markets for the vast majority of our sample period. This is in line with our impulse-response analysis which shows that any shock to stock return quickly disappears most of the time – indicating that the market is generally efficient over the sample period. However, a little evidence for inefficient markets is also discovered. They are, during: (i) the longest recession defined by NBER (1873-1879); (ii) the 1902-1904 recession; (iii) the New Deal era; and (iv) just after the very severe 1957-1958 recession. These results suggest that the deviation from efficient markets occurs in extraordinary times, but not during times of panic (or bubble). This, in turn, raises another question as to whether the market is inefficient due to irrationality or the market is efficient but the individuals' stochastic discount factor changes dramatically. Yet, our results reveal that a sudden alteration of the consumption path does not change the stochastic discount factor abruptly during such time periods. Hence, market inefficiency is likely due to market participants' irrational behavior or shifts in their behavior (the degree of risk aversion or the subjective discount factor), which is partly in accordance with a view of behavioral finance. While this last point requires further investigation, we believe that our approach -- introducing the concept of the degree of market efficiency and allowing for the possibility of the evolving market and -- will provide researchers with a more in-depth view of market efficiency.

\section*{Acknowledgments}

We would like to thank the co-editor, David Peel, two anonymous referees, Kohei Aono, Jun Ma, Colin McKenzie, Taisuke Otsu, Toshiaki Watanabe, Tomoyoshi Yabu, Makoto Yano and participants at the 87th annual conference of the Western Economic Association International in San Francisco for their helpful comments and suggestions. We would also like to thank the Japan Society for the Promotion of Science for their financial assistance provided through the Grant in Aid for Scientific Research Nos. 26380397 (Mikio Ito), and 24530364/15K03542 (Akihiko Noda). All data and programs used for this paper are available upon request.

\setcounter{table}{0}
\setcounter{figure}{0}
\setcounter{equation}{0}
\renewcommand{\theequation}{\arabic{equation}}
\renewcommand{\thetable}{\arabic{table}}
\renewcommand{\thefigure}{\arabic{figure}}
\renewcommand{\thesubsection}{\arabic{subsection}}

\clearpage

\begin{figure}[bp]
 \caption{The Returns on S\&P500}
 \label{ev_fig1}
 \begin{center}
 \includegraphics[scale=0.8]{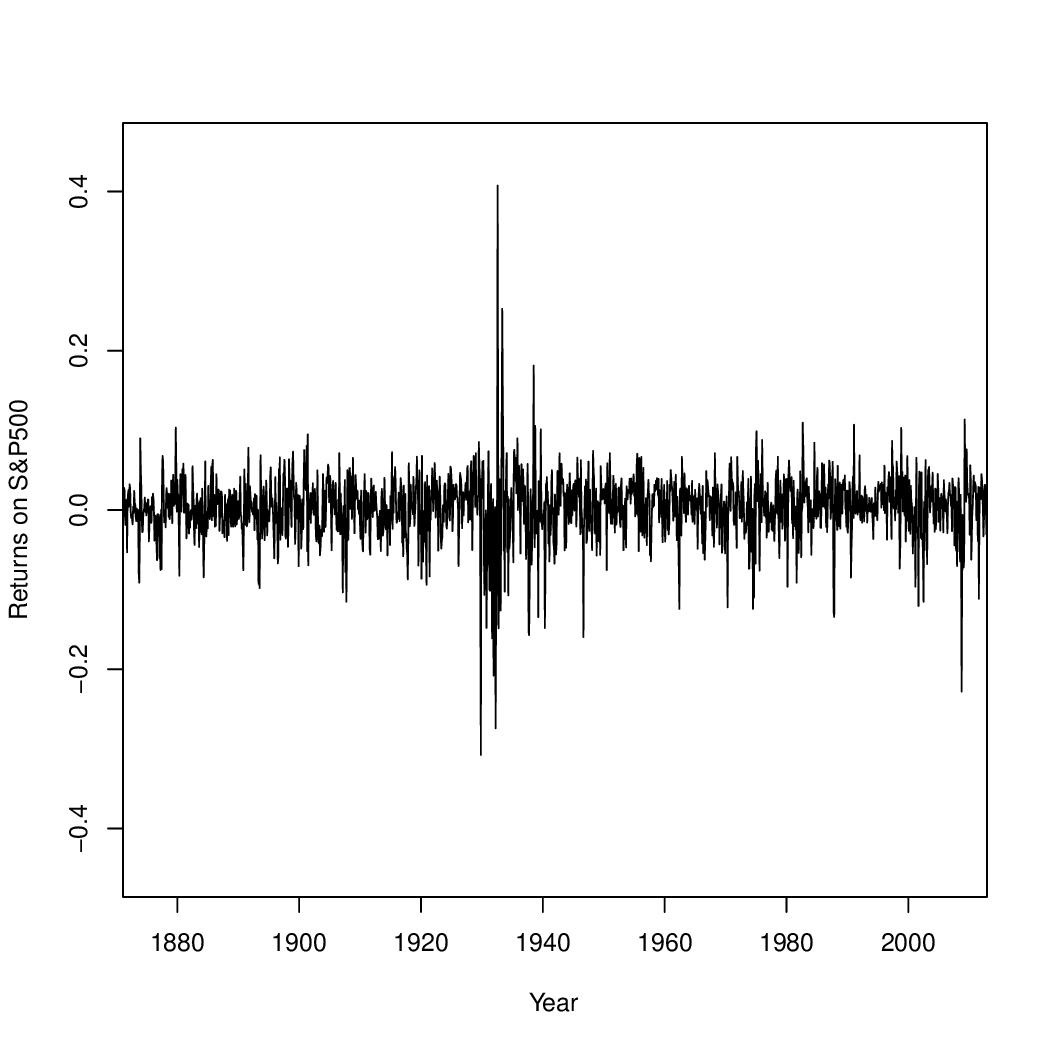}
 \end{center}
\end{figure}

\clearpage

\begin{figure}[bp]
 \caption{Optimal Weights for the Smoother}
 \label{ev_fig2}
 \begin{center}
 \includegraphics[scale=0.8]{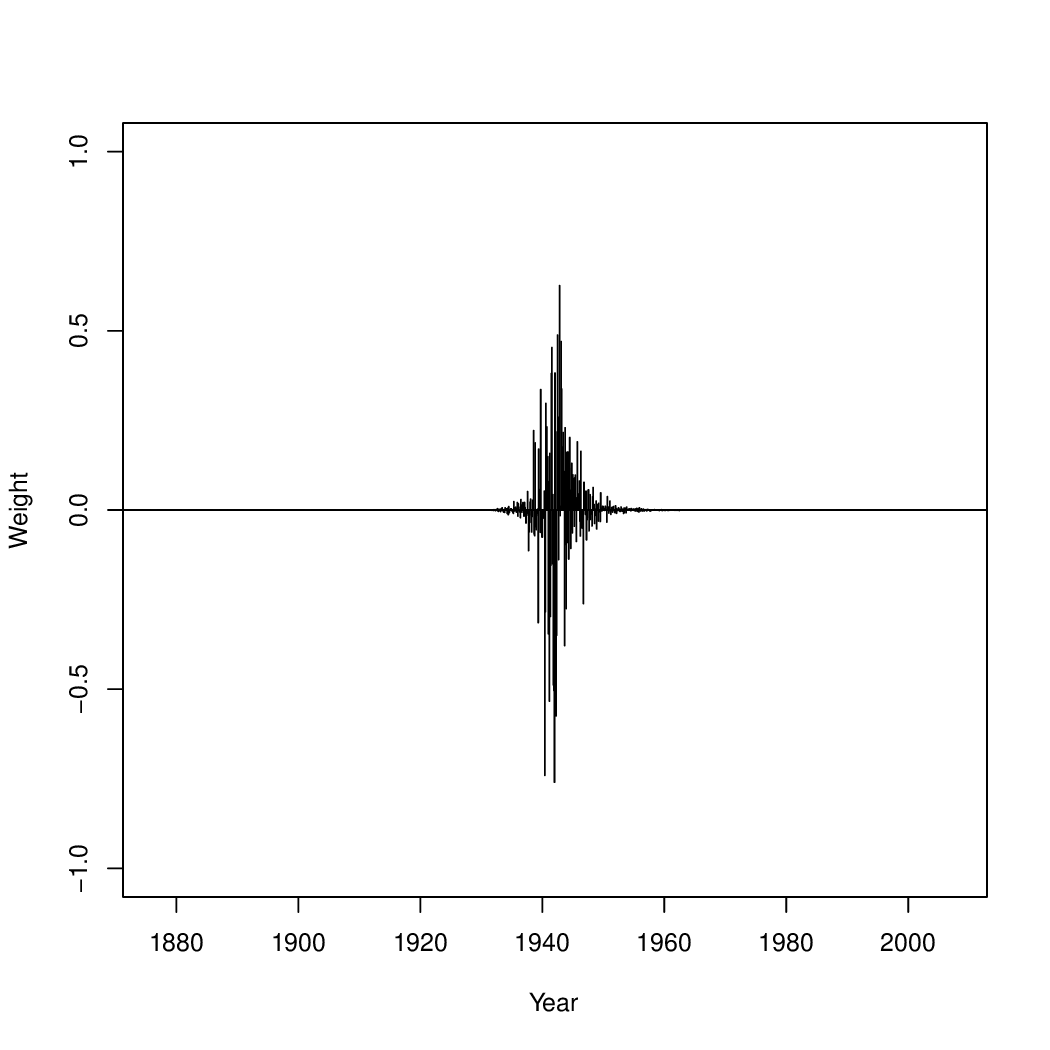}
 \end{center}
\end{figure}

\clearpage

\begin{table}[bp]
\caption{Descriptive Statistics and Unit Root Test}
\label{ev_table1}
\begin{center}
\begin{tabular}{cccccccccccc} \hline\hline
 & Mean & SD & Min & Max & & ADF-GLS & Lag & $\hat\psi$ & &
 $\mathcal{N}$ \\\cline{2-5}\cline{7-9}\cline{11-11}
 & $0.0034$ & $0.0410$ & $-0.3075$ & $0.4075$ & & $-30.1543$ & $0$
			     & $0.3033$ & & $1703$ & \\\hline\hline
\end{tabular}
\vspace*{5pt}
{
\begin{minipage}{400pt}
\footnotesize
{\underline{Notes:}}
\begin{itemize}
\item[(1)] ``ADF-GLS'' denotes the ADF-GLS test statistics, ``Lag''
	   denotes the lag order selected by the MBIC, and
	   ``$\hat\psi$'' denotes the coefficients vector in the GLS
	   detrended series (see equation (6) in \citet{ng2001lls}).
\item[(2)]  In computing the ADF-GLS test, a model with a time trend and
	    a constant is assumed. The critical value at the 1\%
	    significance level for the ADF-GLS test is ``$-3.42$.''
\item[(3)] ``$\mathcal{N}$'' denotes the number of observations.
\item[(4)] R version 3.2.2 was used to compute the statistics.
\end{itemize}
\end{minipage}}%
\end{center}%
\end{table}%

\clearpage

\begin{table}[bp]
\caption{Preliminary Estimation and Parameter Constancy Test}
\label{ev_table2}
\begin{center}
\begin{tabular}{ccccccccc} \hline\hline
 & $Constant$ & $R_{t-1}$ & $R_{t-2}$ & & ${\bar R}^2$ & &
 $L_C$ & \\\cline{2-4}\cline{6-6}\cline{8-8}
 & $0.0026$ & $0.3089$ & $-0.0808$ & & \multirow{2}*{0.0860}
			 & & \multirow{2}*{53.4309} & \\
 & $[0.0010]$ & $[0.0281]$ & $[0.0308]$ & & & & & \\ \hline\hline
\end{tabular}
\vspace*{5pt}
{
\begin{minipage}{290pt}
\footnotesize
{\underline{Notes:}}
\begin{itemize}
\item[(1)] ``$R_{t-1}$,'' ``$R_{t-2}$,'' ``$\bar{R}^2$,'' and ``$L_C$''
 	   denote the AR(1) estimate, the AR(2) estimate, the adjusted
 	   $R^2$, and \citetapos{hansen1992a} joint $L$ statistic with 
           variance, respectively.
\item[(2)] \citetapos{newey1987sps} robust standard errors are in brackets.
\item[(3)] R version 3.2.2 was used to compute the estimates.
\end{itemize}
\end{minipage}}%
\end{center}%
\end{table}%

\clearpage

\begin{figure}[bp]
 \caption{Non-Bayesian TV-AR Estimation}
 \label{ev_fig3}
 \begin{center}
 \includegraphics[scale=0.8]{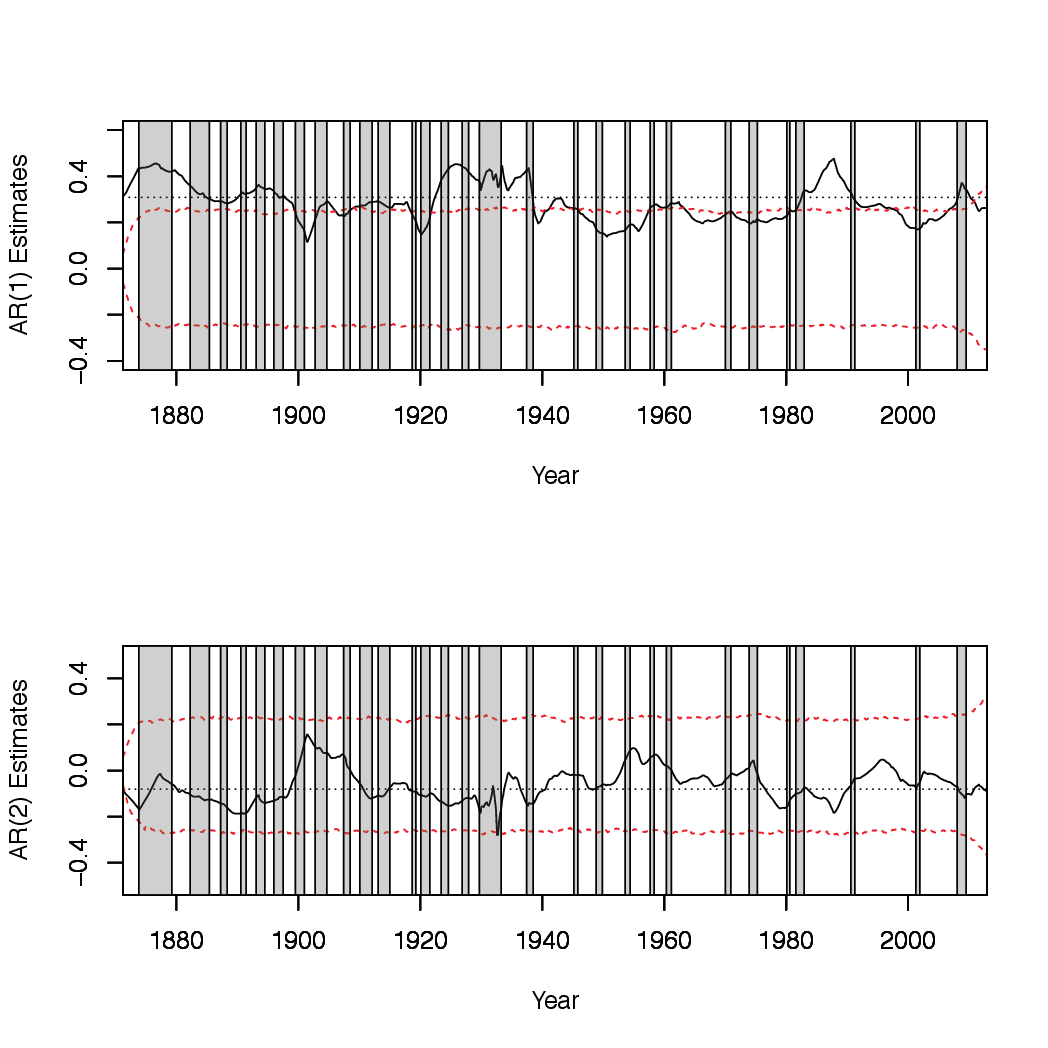}
\vspace*{3pt}
{
\begin{minipage}{350pt}
\footnotesize
\underline{Notes}:
\begin{itemize}
\item[(1)] The dashed red lines represent the 99\% confidence bands of 
           the TV-AR estimates in case of efficient market.
\item[(2)] We run 5000 times Monte Carlo sampling to calculate the 
           confidence bands. 
\item[(3)] The dotted lines also represent the TV-AR estimates with whole 
           sample, respectively.
\item[(4)] The shaded areas are recessions as defined by the NBER.
\item[(5)] R version 3.2.2 was used to compute the estimates.
\end{itemize}
\end{minipage}}%
 \end{center}
\end{figure}

\clearpage

\begin{figure}[bp]
 \caption{Time-Varying Impulse Responses}
 \label{ev_fig4}
 \begin{center}
 \includegraphics[scale=0.8]{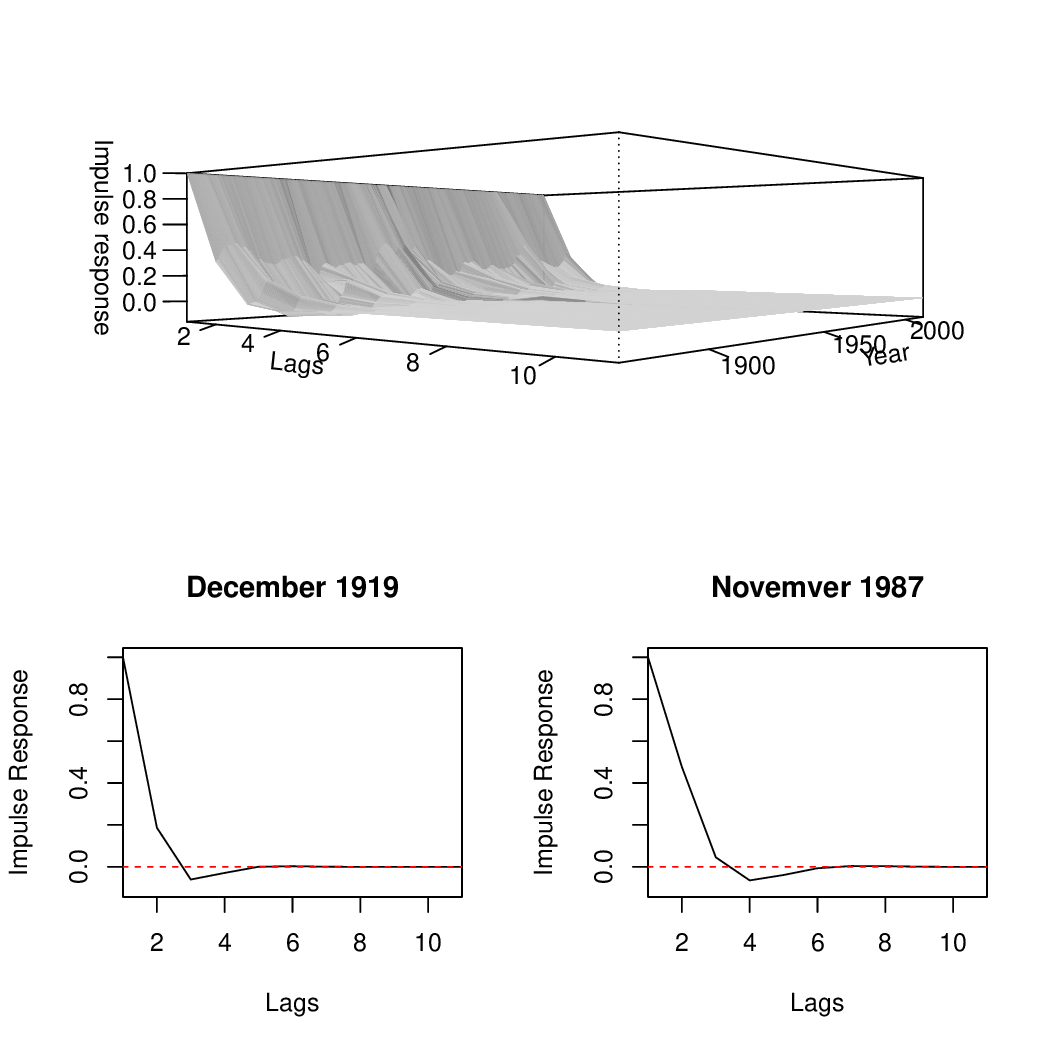}
 \end{center}
\end{figure}

\clearpage

\begin{figure}[bp]
 \caption{Time-Varying Long-Run Multiplier}
 \label{ev_fig5}
 \begin{center}
 \includegraphics[scale=0.8]{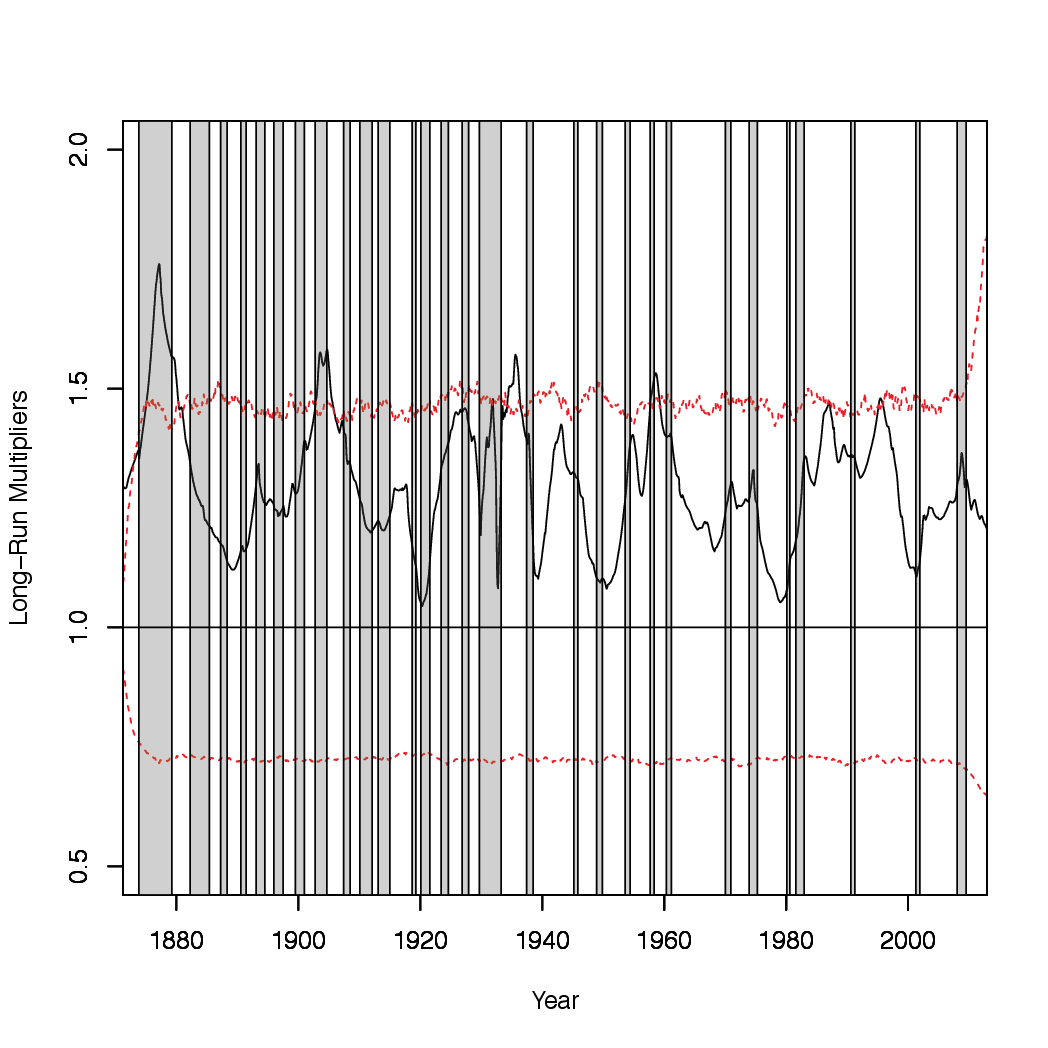}
\vspace*{3pt}
{
\begin{minipage}{350pt}
\footnotesize
\underline{Notes}:
\begin{itemize}
\item[(1)] The dashed red lines represent the 99\% confidence bands of 
           the time-varying long-run multiplier in case of efficient 
           market.
\item[(2)] We run 5000 times Monte Carlo sampling to calculate the 
           confidence bands.
\item[(3)] The shaded areas are recessions as defined by the NBER.
\item[(4)] R version 3.2.2 was used to compute the estimates.
\end{itemize}
\end{minipage}}%
 \end{center}
\end{figure}

\clearpage

\begin{figure}[bp]
 \caption{Power Spectrum Analysis}
 \label{ev_fig6}
 \begin{center}
 \includegraphics[scale=0.8]{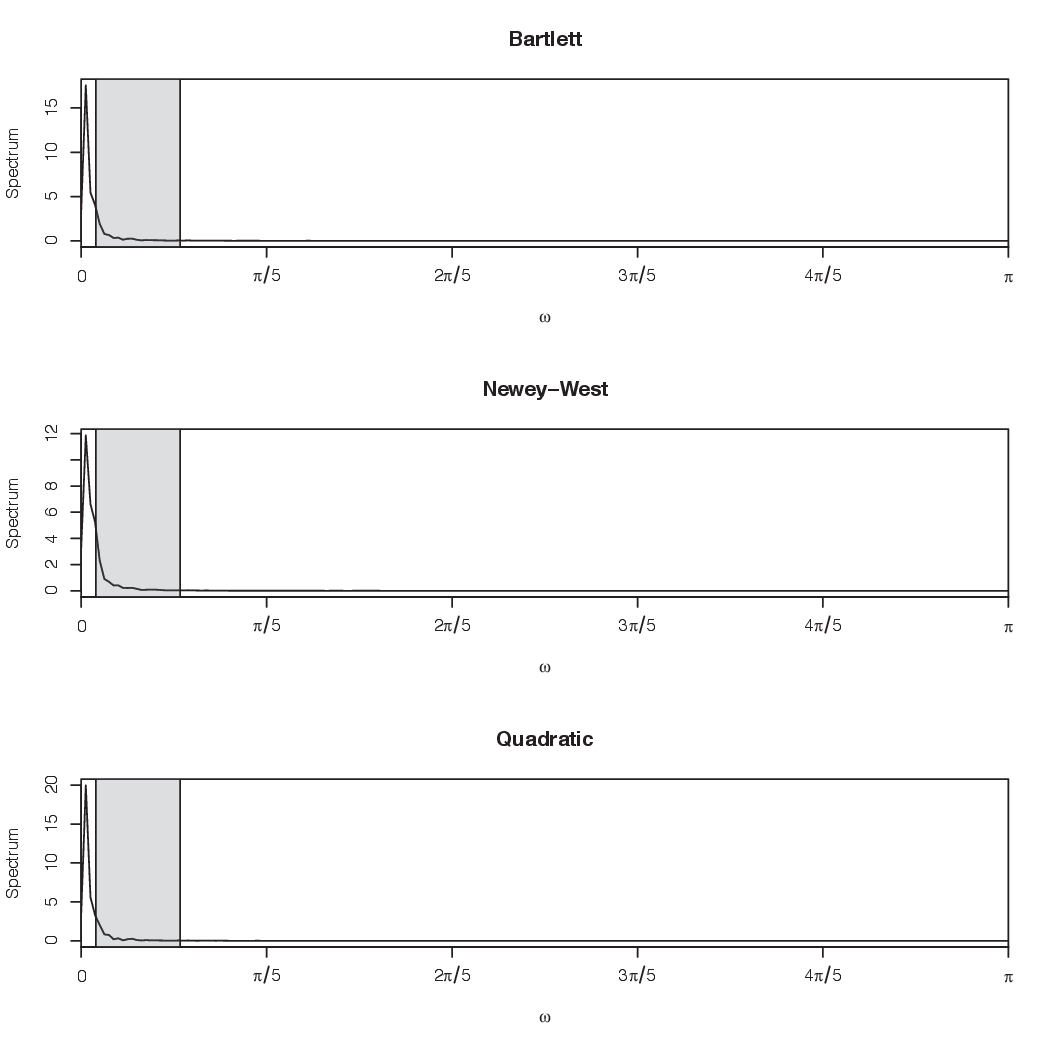}
\vspace*{3pt}
{
\begin{minipage}{350pt}
\footnotesize
\underline{Notes}:
\begin{itemize}
\item[(1)] ``Bartlett,'' ``Newey-West,'' and ``Quadratic'' denote the 
           type of the window kernel to estimate the power spectrums, 
           respectively.
\item[(2)] The shaded area corresponds to the standard business cycle 
           frequencies (72-384 months).
\item[(3)] R version 3.2.2 was used to compute the statistics.
\end{itemize}
\end{minipage}}%
 \end{center}
\end{figure}

\clearpage

\begin{figure}[bp]
 \caption{Industrial Production Growth Rate}
 \label{ev_fig7}
 \begin{center}
 \includegraphics[scale=0.8]{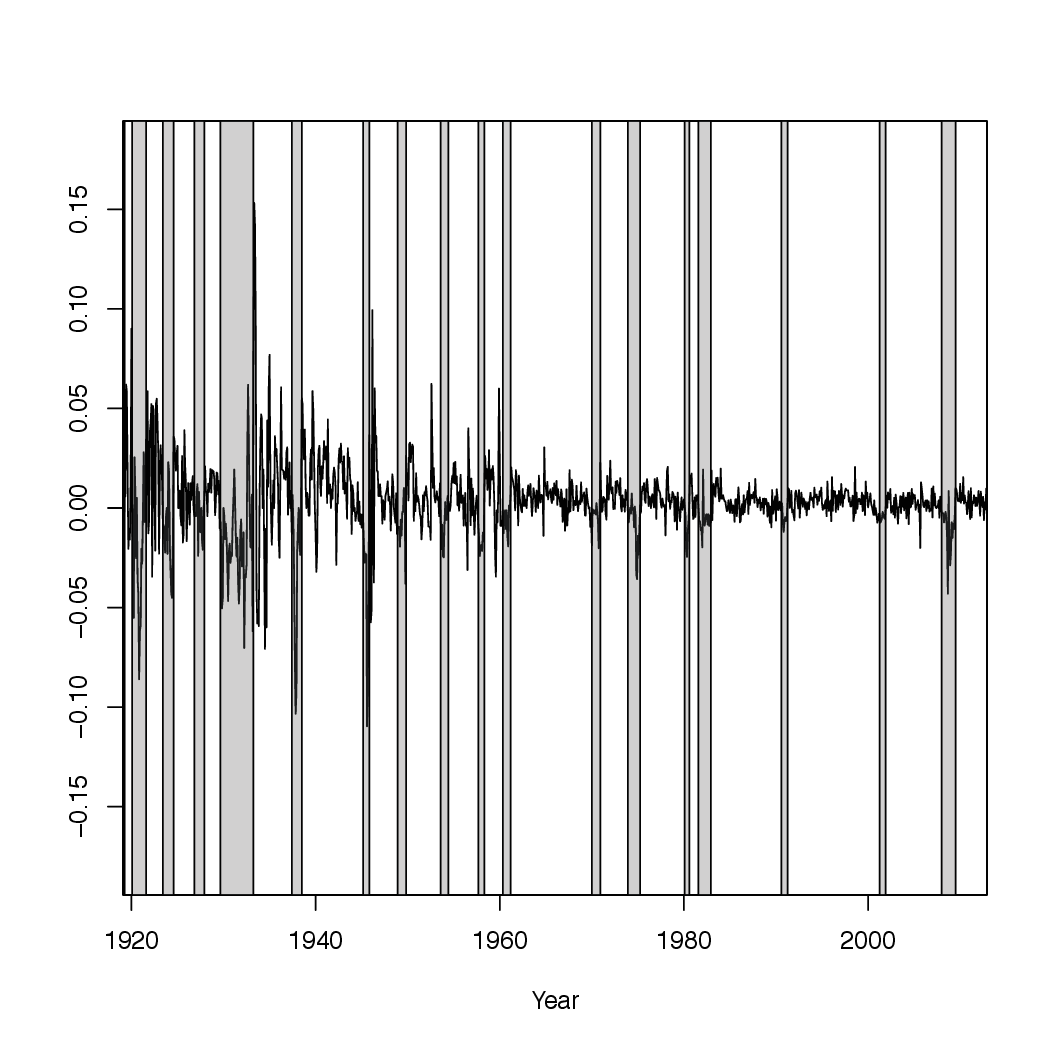}
\vspace*{3pt}
{
\begin{minipage}{350pt}
\footnotesize
\underline{Notes}:
\begin{itemize}
\item[(1)] The shaded areas are recessions as defined by the NBER.
\item[(2)] R version 3.2.2 was used to compute the statistics.
\end{itemize}
\end{minipage}}%
 \end{center}
\end{figure}


\begin{thebibliography}{54}
\providecommand{\natexlab}[1]{#1}

\bibitem[{Andrews(1991)}]{andrews1991hac}
Andrews, D. W.~K. (1991) Heteroskedasticity and autocorrelation consistent
  covariance matrix estimation, \textit{Econometrica}, \textbf{59}, 817--858.

\bibitem[{Andrews(1993)}]{andrews1993emu}
Andrews, D. W.~K. (1993) Exactly median-unbiased estimation of first order
  autoregressive/unit root models, \textit{Econometrica}, \textbf{61},
  139--165.

\bibitem[{Ariel(1987)}]{ariel1987ame}
Ariel, R.~A. (1987) A monthly effect in stock returns, \textit{Journal of
  Financial Economics}, \textbf{18}, 161--174.

\bibitem[{Ariel(1990)}]{ariel1990hsr}
Ariel, R.~A. (1990) High stock returns before holidays: Existence and evidence
  on possible causes, \textit{Journal of Finance}, \textbf{45}, 1611--1626.

\bibitem[{Banz(1981)}]{banz1981trr}
Banz, R.~W. (1981) The relationship between return and market value of common
  stocks, \textit{Journal of Financial Economics}, \textbf{9}, 3--18.

\bibitem[{Baxter and King(1999)}]{baxter1999mbc}
Baxter, M. and King, R.~G. (1999) Measuring business cycles: Approximate
  band-pass filters for economic time series, \textit{Review of Economics and
  Statistics}, \textbf{81}, 575--593.

\bibitem[{Beveridge and Nelson(1981)}]{beveridge1981nad}
Beveridge, S. and Nelson, C.~R. (1981) A new approach to decomposition of
  economic time series into permanent and transitory components with particular
  attention to measurement of the `business cycle', \textit{Journal of Monetary
  Economics}, \textbf{7}, 151--174.

\bibitem[{Blume(1970)}]{blume1970pta}
Blume, M.~E. (1970) Portfolio theory: A step toward its practical application,
  \textit{Journal of Business}, \textbf{43}, 152--173.

\bibitem[{Brockwell and Davis(1991)}]{brockwell1991tst}
Brockwell, P.~J. and Davis, R.~A. (1991) \textit{Time Series: Theory and
  Methods}, Springer, 2nd edn.

\bibitem[{Chan and Chen(1991)}]{chan1991src}
Chan, K.~C. and Chen, N. (1991) Structural and return characteristics of small
  and large firms, \textit{Journal of Finance}, \textbf{46}, 1467--1484.

\bibitem[{Cochrane(2005)}]{cochrane2005ap}
Cochrane, J.~H. (2005) \textit{Asset Pricing}, Princeton University Press.

\bibitem[{Conrad and Kaul(1988)}]{conrad1988tve}
Conrad, J. and Kaul, G. (1988) Time-variation in expected returns,
  \textit{Journal of Business}, \textbf{61}, 409--425.

\bibitem[{Elliott \textit{et~al.}(1996)Elliott, Rothenberg and
  Stock}]{elliott1996eta}
Elliott, G., Rothenberg, T.~J. and Stock, J.~H. (1996) Efficient tests for an
  autoregressive unit root, \textit{Econometrica}, \textbf{64}, 813--836.

\bibitem[{Emerson \textit{et~al.}(1997)Emerson, Hall and
  Zalewska-Mitura}]{emerson1997eme}
Emerson, R., Hall, S.~G. and Zalewska-Mitura, A. (1997) Evolving market
  efficiency with an application to some bulgarian shares, \textit{Economics of
  Planning}, \textbf{30}, 75--90.

\bibitem[{Fama(1965)}]{fama1965bsm}
Fama, E.~F. (1965) The behavior of stock-market prices, \textit{Journal of
  Business}, \textbf{38}, 34--105.

\bibitem[{Fama(1970)}]{fama1970ecm}
Fama, E.~F. (1970) Efficient capital markets: A review of theory and empirical
  work, \textit{Journal of Finance}, \textbf{25}, 383--417.

\bibitem[{Fama(1991)}]{fama1991ecm}
Fama, E.~F. (1991) Efficient capital markets: {I\hspace{-.1em}I},
  \textit{Journal of Finance}, \textbf{46}, 1575--1617.

\bibitem[{Fama(1998)}]{fama1998eml}
Fama, E.~F. (1998) Market eifficiency, long-term returns, and behavioral
  finance, \textit{Journal of Financial Economics}, \textbf{49}, 283--306.

\bibitem[{Fama and Blume(1966)}]{fama1966fra}
Fama, E.~F. and Blume, M.~E. (1966) Filter rules and stock-market trading,
  \textit{Journal of Business}, \textbf{39}, 226--241.

\bibitem[{Fama \textit{et~al.}(1969)Fama, Fisher, Jensen and
  Roll}]{fama1969asp}
Fama, E.~F., Fisher, L., Jensen, M.~C. and Roll, R. (1969) The adjustment of
  stock prices to new information, \textit{International Economic Review},
  \textbf{10}, 1--21.

\bibitem[{Fama and French(1992)}]{fama1992tcs}
Fama, E.~F. and French, K.~R. (1992) The cross section of expected stock
  returns, \textit{Journal of Finance}, \textbf{47}, 427--465.

\bibitem[{Gu and Finnerty(2002)}]{gu2002eme}
Gu, A.~Y. and Finnerty, J. (2002) The evolution of market efficiency: 103 years
  daily data of the dow, \textit{Review of Quantitative Finance and
  Accounting}, \textbf{18}, 219--237.

\bibitem[{Hamilton(1994)}]{hamilton1994tsa}
Hamilton, J.~D. (1994) \textit{Time Series Analysis}, Princeton University
  Press.

\bibitem[{Hansen(1992)}]{hansen1992a}
Hansen, B.~E. (1992) Testing for parameter instability in linear models,
  \textit{Journal of Policy Modeling}, \textbf{14}, 517--533.

\bibitem[{Harris(1986)}]{harris1986atd}
Harris, L. (1986) A transaction data study of weekly and intradaily patterns in
  stock returns, \textit{Journal of Financial Economics}, \textbf{16}, 99--117.

\bibitem[{Ito(2007)}]{ito2007nme}
Ito, M. (2007) A new method for estimating economic models with general
  time-varying structures, keio Economic Society Discussion Paper Series, KESDP
  07-8.

\bibitem[{Ito and Sugiyama(2009)}]{ito2009mdt}
Ito, M. and Sugiyama, S. (2009) Measuring the degree of time varying market
  inefficiency, \textit{Economics Letters}, \textbf{103}, 62--64.

\bibitem[{Jensen(1968)}]{jensen1968tpm}
Jensen, M.~C. (1968) The performance of mutual funds in the period 1945-1964,
  \textit{Journal of Finance}, \textbf{23}, 389--416.

\bibitem[{Keim(1983)}]{keim1983sra}
Keim, D.~B. (1983) Size-related anomalies and stock return seasonality: Further
  empirical evidence, \textit{Journal of Financial Economics}, \textbf{12},
  13--32.

\bibitem[{Keim(1989)}]{keim1989tpb}
Keim, D.~B. (1989) Trading patterns, bid-ask spreads, and estimated security
  returns: The case of common stocks at calendar turning points,
  \textit{Journal of Financial Economics}, \textbf{25}, 75--97.

\bibitem[{Kim \textit{et~al.}(2011)Kim, Shamsuddin and Lim}]{kim2011srp}
Kim, J.~H., Shamsuddin, A. and Lim, K.~P. (2011) Stock return predictability
  and the adaptive markets hypothesis: Evidence from century-long u.s. data,
  \textit{Journal of Empirical Finance}, \textbf{18}, 868--879.

\bibitem[{King \textit{et~al.}(1988)King, Plosser and Rebelo}]{king1988pgb}
King, R.~G., Plosser, C.~I. and Rebelo, S.~T. (1988) Production, growth and
  business cycles: {I}. the basic neoclassical model, \textit{Journal of
  Monetary Economics}, \textbf{21}, 195--232.

\bibitem[{Lim and Brooks(2011)}]{lim2011esm}
Lim, K.~P. and Brooks, R. (2011) The evolution of stock market efficiency over
  time: a survey of the empirical literature, \textit{Journal of Economic
  Surveys}, \textbf{25}, 69--108.

\bibitem[{Lim \textit{et~al.}(2013)Lim, Luo and Kim}]{lim2013aus}
Lim, K.~P., Luo, W. and Kim, J.~H. (2013) Are {US} stock index returns
  predictable? evidence from automatic autocorrelation-based tests,
  \textit{Applied Economics}, \textbf{45}, 953--962.

\bibitem[{Ljung and Box(1978)}]{ljung1978oml}
Ljung, G.~M. and Box, G. E.~P. (1978) On a measure of lack of fit in time
  series models, \textit{Biometrika}, \textbf{65}, 297--303.

\bibitem[{Lo(2004)}]{lo2004amh}
Lo, A.~W. (2004) The adaptive markets hypothesis: Market efficiency from an
  evolutionary perspective, \textit{Journal of Portfolio Management},
  \textbf{30}, 15--29.

\bibitem[{Lo(2005)}]{lo2005rem}
Lo, A.~W. (2005) Reconciling efficient markets with behavioral finance: The
  adaptive markets hypothesis, \textit{Journal of Investment Consulting},
  \textbf{7}, 21--44.

\bibitem[{L{\"u}tkepohl(2005)}]{lutkepohl2005nim}
L{\"u}tkepohl, H. (2005) \textit{New Introduction to Multiple Time Series
  Analysis}, Springer.

\bibitem[{Maddala and Kim(1998)}]{maddala1998urc}
Maddala, G.~S. and Kim, I. (1998) \textit{Unit Roots, Cointegration, and
  Structural Change}, Cambridge University Press.

\bibitem[{Malkiel(2003)}]{malkiel2003emh}
Malkiel, B.~G. (2003) The efficient market hypothesis and its critics,
  \textit{Journal of Economic Perspectives}, \textbf{17}, 59--82.

\bibitem[{Malkiel \textit{et~al.}(2005)Malkiel, Mullainathan and
  Stangle}]{malkiel2005meb}
Malkiel, B.~G., Mullainathan, S. and Stangle, B.~E. (2005) Market efficiency
  versus behavioral finance, \textit{Journal of Applied Corporate Finance},
  \textbf{17}, 124--136.

\bibitem[{Morley \textit{et~al.}(2003)Morley, Nelson and Zivot}]{morely2003wbn}
Morley, J.~C., Nelson, C.~R. and Zivot, E. (2003) Why are the beveridge-nelson
  and unobserved-components decompositions of gdp so different?, \textit{Review
  of Economics and Statistics}, \textbf{85}, 235--243.

\bibitem[{Newey and West(1987)}]{newey1987sps}
Newey, W.~K. and West, K.~D. (1987) A simple, positive semi-definite,
  heteroskedasticity and autocorrelation consistent covariance matrix,
  \textit{Econometrica}, \textbf{55}, 703--708.

\bibitem[{Newey and West(1994)}]{newey1994als}
Newey, W.~K. and West, K.~D. (1994) Automatic lag selection in covariance
  matrix estimation, \textit{Review of Economic Studies}, \textbf{61},
  631--653.

\bibitem[{Ng and Perron(2001)}]{ng2001lls}
Ng, S. and Perron, P. (2001) Lag length selection and the construction of unit
  root tests with good size and power, \textit{Econometrica}, \textbf{69},
  1519--1554.

\bibitem[{Nyblom(1989)}]{nyblom1989tcp}
Nyblom, J. (1989) Testing for the constancy of parameters over time,
  \textit{Journal of the American Statistical Association}, \textbf{84},
  223--230.

\bibitem[{Paige and Saunders(1977)}]{paige1977lse}
Paige, C.~C. and Saunders, M.~A. (1977) Least squares estimation of discrete
  linear dynamic systems using orthogonal transformations, \textit{SIAM Journal
  on Numerical Analysis}, \textbf{14}, 180--193.

\bibitem[{Perron(1989)}]{perron1989tgc}
Perron, P. (1989) The great crash, the oil price shock, and the unit root
  hypothesis, \textit{Econometrica}, \textbf{57}, 1361--1401.

\bibitem[{Perron and Wada(2009)}]{perron2009ltb}
Perron, P. and Wada, T. (2009) Let's take a break: Trends and cycles in us real
  gdp, \textit{Journal of Monetary Economics}, \textbf{56}, 749--765.

\bibitem[{Sargent(1987)}]{sargent1987mt}
Sargent, T. (1987) \textit{Macroeconomic Theory}, Academic Press, 2nd edition
  edn.

\bibitem[{Schwarz(1978)}]{schwarz1978edm}
Schwarz, G. (1978) Estimating the dimension of a model, \textit{Annals of
  Statistics}, \textbf{6}, 461--464.

\bibitem[{Schwert(2003)}]{schwert2003ame}
Schwert, G.~W. (2003) Anomalies and market efficiency, in \textit{Handbook of
  the Economics of Finance} (Eds.) G.~M. Constantinides, M.~Harris and R.~M.
  Stulz, North-Holland, chap.~15, pp. 937--972.

\bibitem[{Yen and Lee(2008)}]{yen2008wmh}
Yen, G. and Lee, C. (2008) Efficient market hypothesis ({EMH}): Past, present
  and future, \textit{Review of Pacific Basin Financial Markets and Policies},
  \textbf{11}, 305--329.

\bibitem[{Zalewska-Mitura and Hall(1999)}]{zalewska1999efs}
Zalewska-Mitura, A. and Hall, S.~G. (1999) Examining the first stages of market
  performance: a test for evolving market efficiency, \textit{Economics
  Letters}, \textbf{64}, 1--12.

\end{thebibliography}
\end{document}